\documentclass[12pt]{iopart}          

\usepackage{iopams} 
\usepackage{graphicx}% Include figure files
\usepackage{longtable}
\usepackage{array}
\usepackage{dcolumn}% Align table columns on decimal point
\usepackage{bm}% bold math
\usepackage{float}
\usepackage[british]{babel}
\usepackage{multirow}
\usepackage{caption}
\usepackage{siunitx}

\usepackage{lipsum}  
\usepackage{rotating}
\usepackage{color}
\usepackage{xcolor}
\usepackage{hyperref}
\hypersetup
{
pdftitle="Deep Learning Based Event Reconstruction for Cyclotron Radiation Emission Spectroscopy",
pdfauthor="Luis Saldana",
pdftoolbar=true, 
pdfmenubar=true, 
pdfstartview={FitH},
colorlinks=true, 
pdfpagemode=UseNone, 
pdfpagelayout=SinglePage, 
pdffitwindow=true, 
linkcolor=blue,
citecolor=blue,
urlcolor=blue 
}

\usepackage[utf8x]{inputenc}
\usepackage[T1]{fontenc}

\usepackage{footmisc}
\newcommand*\diff{\mathop{}\!\mathrm{d}} %% for differential
\usepackage[font=small]{caption}
\usepackage{amsmath}
\usepackage{siunitx}
\DeclareUnicodeCharacter{2212}{\ensuremath{-}}

\begin{document}

\title[Deep Learning Based Event Reconstruction for CRES]{Deep Learning Based Event Reconstruction for Cyclotron Radiation Emission Spectroscopy}

\author{
A.~Ashtari~Esfahani$^1$,
S.~Böser$^2$,
N.~Buzinsky$^3$,
M.~C.~Carmona-Benitez$^4$,
R.~Cervantes$^1$,
C.~Claessens$^1$,
L.~de~Viveiros$^4$,
M.~Fertl$^2$,
J.~A.~Formaggio$^3$,
J.~K.~Gaison$^5$,
L.~Gladstone$^6$,
M.~Grando$^5$,
M.~Guigue$^5$,
J.~Hartse$^1$,
K.~M.~Heeger$^7$,
X.~Huyan$^5$,
A.~M.~Jones$^5$,
K.~Kazkaz$^8$,
M.~Li$^3$,
A.~Lindman$^2$,
A.~Marsteller$^1$,
C.~Matthé$^2$,
R.~Mohiuddin$^6$,
B.~Monreal$^6$,
E.~C.~Morrison$^5$,
R.~Mueller$^4$,
J.~A.~Nikkel$^7$,
E.~Novitski$^1$,
N.~S.~Oblath$^5$,
J.~I.~Peña$^3$,
W.~Pettus$^9$,
R.~Reimann$^2$,
R.~G.~H.~Robertson$^1$,
L.~Saldaña$^{7,\mathsection}$,
M.~Schram$^5$,
P.~L.~Slocum$^7$,
J.~Stachurska$^3$,
Y.-H.~Sun$^6$,
P.~T.~Surukuchi$^7$,
A.~B.~Telles$^7$,
F.~Thomas$^2$,
M.~Thomas$^5$,
L.~A.~Thorne$^2$,
T.~Thümmler$^{10}$,
L.~Tvrznikova$^8$,
W.~Van De Pontseele$^3$,
B.~A.~VanDevender$^{1,5}$,
T.~E.~Weiss$^7$,
T.~Wendler$^4$,
E.~Zayas$^3$,
A.~Ziegler$^4$
}

%L.~Salda\~na$^{1\mathsection}$,
%R.~Mohiuddin$^2$,
%Y.-H.~Sun$^2$,
%P.~T.~Surukuchi$^1$
%W.~Van De Pontseele$^3$

\ead{$^{\mathsection}$luis.saldana@yale.edu}
\address{$^1$ Center for Experimental Nuclear Physics and Astrophysics and Department of Physics, University of Washington, Seattle, WA 98195, USA}
\address{$^2$ Institute for Physics, Johannes Gutenberg University Mainz, Mainz, Germany 55099}
\address{$^3$ Laboratory for Nuclear Science, Massachusetts Institute of Technology, Cambridge, MA 02139, USA}
\address{$^4$ Department of Physics, Pennsylvania State University, University Park, PA 16802, USA}
\address{$^5$ Pacific Northwest National Laboratory, Richland, WA 99354, USA}
\address{$^6$ Department of Physics, Case Western Reserve University, Cleveland, OH 44106, USA}
\address{$^7$ Wright Laboratory, Department of Physics, Yale University, New Haven, CT 06520, USA}
\address{$^8$ Nuclear and Chemical Sciences, Lawrence Livermore National Laboratory, Livermore, CA, USA 94550}
\address{$^{9}$ Center for Exploration of Energy and Matter and Department of Physics, Indiana University, Bloomington, IN 47405, USA}
\address{$^{10}$ Institute for Astroparticle Physics, Karlsruhe Institute of Technology, Karlsruhe, Germany 76021}

%% Abstract
\begin{abstract}
{
The objective of the Cyclotron Radiation Emission Spectroscopy (CRES) technology is to build precise particle energy spectra. This is achieved by identifying the start frequencies of charged particle trajectories which, when exposed to an external magnetic field, leave semi-linear profiles (called tracks) in the time-frequency plane. Due to the need for excellent instrumental energy resolution in application, highly efficient and accurate track reconstruction methods are desired. Deep learning convolutional neural networks (CNNs) - particularly suited to deal with information-sparse data and which offer precise foreground localization - may be utilized to extract track properties from measured CRES signals (called events) with relative computational ease. In this work, we develop a novel machine learning based model which operates a CNN and a support vector machine in tandem to perform this reconstruction. A primary application of our method is shown on simulated CRES signals which mimic those of the Project 8 experiment - a novel effort to extract the unknown absolute neutrino mass value from a precise measurement of tritium $\beta^-$-decay energy spectrum. When compared to a point-clustering based technique used as a baseline, we show a relative gain of 24.1\% in event reconstruction efficiency and comparable performance in accuracy of track parameter reconstruction.
}
\end{abstract}

\newline
\submitto{Machine Learning: Science and Technology}
%\maketitle
%\tableofcontents

%% Main Body

%%%%%%%%%%%%%%%%%%%% Chapter 1: Introduction %%%%%%%%%%%%%%%%%%%%
%\newpage
\section{Introduction}
\label{sc:intro}

Over the past ten years the use of convolutional neural networks (CNNs) as a machine learning (ML) method has gained considerable attention in the high energy physics community for tasks such as particle identification, event reconstruction, and anomaly detection \cite{ml_dl_in_particle_phys,Radovic2018,ai_ml_nuclearphys}. 
Due to reduction in number of fully-connected layers and featuring translational and rotational equivariance, CNNs originally designed to mimic a simplified version of the animal visual cortex have become the preferred choice for tackling various computer vision problems \cite{bhatt}. 
Of particular interest to this work is their application to the semantic segmentation of images: assigning a classification label to every pixel via learned inference for the purposes of object reconstruction.
In a CNN, a \emph{convolutional} step aids in highlighting and detecting specific aspects of the image, e.g. corners or edges, which become more abstract with sequential applications of \emph{filters} throughout many layers. 
The task of finding the optimal set of filters which most accurately performs the segmentation is relegated to a supervised learning optimization task. With this approach, hand-engineering of features and hand-tuning of model parameters become obsolete. 

\begin{figure}[ht!]
\begin{center}
\includegraphics[width=0.7\textwidth]{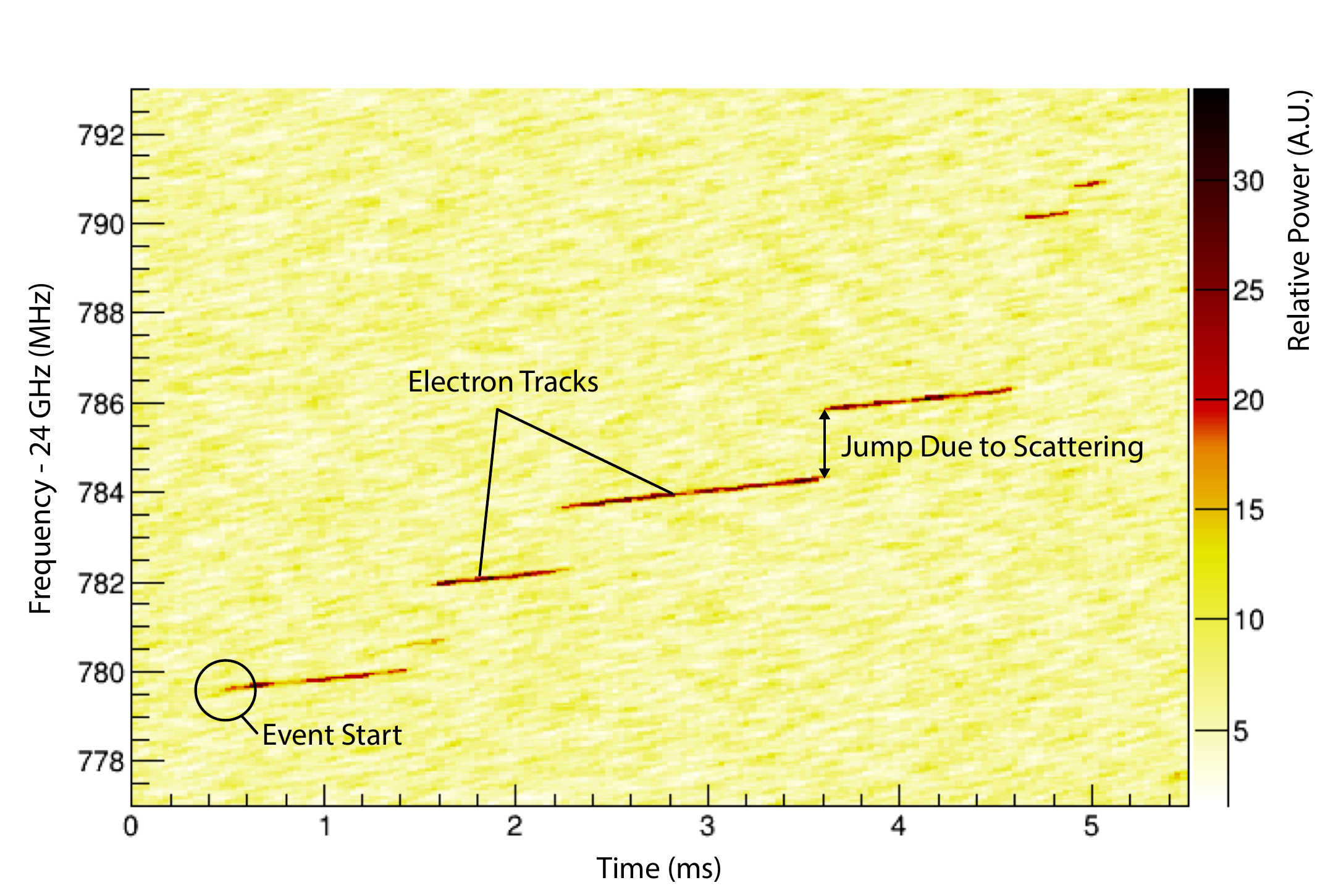}
\caption{A CRES signal as seen in a spectrogram from the Project 8 experiment. Here, a single internal conversion electron from $^{83\text{m}}$Kr makes an event consisting of multiple tracks scattered in frequency due to rapid energy losses from inelastic collisions with residual gases. Radio-frequency thermal noise is shown as yellow-colored pixels of relatively low power. This figure was filtered to make the event features more visible as foreground.  \label{fig:cres_event}}
\end{center}
\end{figure}

In Cyclotron Radiation Emission Spectroscopy (CRES) experiments~\cite{Monreal:2009za,Project8:2014ivu,he6_cres}, the goal is to reconstruct the trajectories of charged particles, which are referred to as tracks.
These tracks profile as narrow traces over the frequency and time plane and are all contained in images called spectrograms - see Figure~\ref{fig:cres_event}.
The CRES procedure is to accurately identify the track start frequencies and, using the cyclotron motion relationship given in Eq.~\eqref{eq:cyclotron_frequency}, precisely reconstruct the energy spectrum of the underlying physical process.
For example, the spectrum of interest could be that of tritium $\beta^-$-decay electrons for absolute neutrino mass measurements \cite{t2_prl} or of $\beta^{\pm}$-particles in nuclear decays of $^6$He and $^{19}$Ne to study chirality-flipping in the weak sector~\cite{he6_cres}.
In the example spectrogram from the Project 8 experiment\footnote{\href{https://www.project8.org/}{https://www.project8.org/}} shown in Figure \ref{fig:cres_event}, a single tritium $\beta^-$-decay electron is manifested as a collection of multiple tracks within an event, inter-spread in frequency by rapid energy losses due to scattering collisions with residual gases in the apparatus \cite{phenom_paper}. Traditionally, CRES tracks are reconstructed by first filtering a spectrogram for bins of high signal-to-noise ratio\footnote{The SNR is defined as the deposited power divided by the average power of all noise bins in the image.} (SNR) and subsequently applying point-clustering techniques which look for pre-configured patterns in the data.
The resulting reconstruction model is made robust by tuning a large number of parameters by hand  \cite{t2_prc}. 
When we consider that the experimental sensitivity to neutrino mass depends on the number of events observed (and therefore reconstructed) \cite{bayesian_paper} and that the statistical variance on the extracted mass is related to the total number of events in the observation window by $\sigma^2_\text{stat}\sim N_\text{tot}$ \cite{Formaggio_2021}, the need for a reconstruction technique which is both efficient and accurate becomes prominent. 

In this work, we  present an alternative approach to CRES signal reconstruction which utilizes the power of ML-based modules, principally the CNN, to rid the need for pre-engineering of features and parameters in the analysis. 
Additionally, we provide a methodology with demonstrated state-of-the-art performance in segmentation tasks over a large number of applications \cite{MO2022}. 
Nevertheless, despite the relatively simple geometric features of CRES tracks, segmentation becomes challenging in the presence of radio-frequency (RF) thermal noise which serves as a background covering more than 99.99\% of all pixels in a single spectrogram on average. 
The immense class-imbalance problem requires a comprehensive model which is not only accurate but also efficiently rejects false positives. 
The latter problem becomes highly prominent in the presence of very short CRES tracks of low SNR, which may be easily mistaken for random correlated noise fluctuations of high SNR and vice versa.

The novel reconstruction model presented in this work is based on a CNN, acting as a track-pixel segmentation step, and a support vector machine (SVM), acting as a track-object false-positive veto, working in tandem.
This marks the first step towards a fully ML-based reconstruction approach for CRES-type experiments. 
In particular, we focus our application to CRES signal profiles as observed in the Project 8 experiment and perform a validation comparison to the existing baseline reconstruction algorithm over simulated data. 
We proceed by first discussing the signal acquisition and baseline event reconstruction used in the Project 8 experiment in Section~\ref{sc:cres_event_reconstruction}. 
Section~\ref{sc:ML} describes the new ML-based reconstruction method developed for CRES events and Section~\ref{sc:simulation} outlines the generation of simulated data for training and validation.
The optimization performed on both reconstruction methods is described in Section~\ref{sc:optimization}.
Section~\ref{sc:results} summarizes the results from both techniques and the improvements in efficiency achieved by the ML approach.

%%%%%%%%%%%%%%%%%%%% Chapter 2: Event Rec in P8 %%%%%%%%%%%%%%%%%%%%

\section{CRES Event Reconstruction}
\label{sc:cres_event_reconstruction}

CRES works by reconstructing energies from the cyclotron radiation emitted by charged particles when subjected to an external magnetic field \cite{Monreal:2009za}. For example, in the case of tritium $\beta^-$-decay ($\text{T} \rightarrow {}^3\text{He}^+ + e^- + \overline{\nu}_e$), the semi-relativistic daughter electrons carry kinetic energy $E_\text{kin}$ and emit cyclotron radiation with frequency
\begin{equation}
\label{eq:cyclotron_frequency}
f_c =  \frac{1}{2\pi}\frac{|e|B}{m_e+E_{\rm{kin}}/c^2}
\end{equation}
where $m_e$ and $e$ are the electron's mass and charge respectively, and $B$ is the magnitude of the applied magnetic field.
This relationship allows for reconstruction of $E_\text{kin}$ via frequency sampling once the magnetic field is known \cite{Monreal:2009za}. 
Due to conservation of energy and momentum, the mass of the daughter neutrino is extracted by precision analysis of the spectrum near the tritium endpoint ($Q\simeq 18.6$ keV), where the effect of the massive neutrino on the spectral shape is maximal \cite{Formaggio_2021}. 

\begin{figure}[ht!]
\begin{center}
\includegraphics[width=0.7\textwidth]{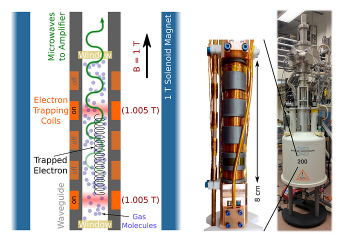}
\caption{In Phase II of the Project 8 experiment, a cylindrical waveguide, which confines both source gas and daughter electrons, is placed inside the bore of an NMR magnet that produces a background $\sim 1$ T field necessary to elicit cyclotron motion. Additional electron trapping coils are wound around the waveguide which aid in creating custom-profile magnetic traps. The emitted cyclotron radiation travels through the upper waveguide window into a chain of low-noise amplifiers (among other components) to be sampled and reconstructed. Schematic and photographs provided by the Project 8 experiment. \label{fig:exp_setup_schematic}}
\end{center}
\end{figure}

For purposes of neutrino mass measurement, the Project 8 experiment sources electrons via $\beta^-$-decay from molecular tritium and via internal conversion from $^{83\text{m}}$Kr; the latter only used for energy calibration. 
Both the source and electrons are confined in a cylindrical waveguide and subjected to an axial $\sim 1$ T field produced by an Nuclear Magnetic Resonance (NMR) magnet. 
Additional current-carrying coils are wound around the waveguide to produce magnetic field gradients of $\mathcal{O}(\text{mT})$ which serve as ``walls'' for axial magnetic trapping.
The resulting emitted cyclotron radiation is directed into a radio-frequency (RF) antenna receiver chain made up of low-noise amplifiers among additional components \cite{t2_prc}. 
A picture of the experiment alongside a schematic of the detector can be seen in Figure \ref{fig:exp_setup_schematic}. 

For an electron emitted near the tritium $\beta^-$-decay endpoint, the cyclotron radiation has a frequency of approximately 26 GHz. 
The raw signal received by the antenna chain is down-mixed, filtered, and sampled by a ROACH2 \cite{roach2} digitizer which performs a Fourier transform to extract frequency information \cite{r2daq_reference}.
The data acquisition software is triggered in the presence of high power bins in the frequency-domain and the original raw time series is subsequently saved to disk for offline event reconstruction.

The time series data from the antennas consists of a CRES signal superimposed onto an RF noise background.
In order to extract the signal information from this, a series of offline short-time Fast Fourier Transform (FFT) are performed which, when stacked, result in a spectrogram describing the evolution of the frequency profile over time (see Figure \ref{fig:cres_event}). 
A generic CRES event consists of one or multiple tracks separated by jumps in frequency, due to energy losses from scattering off residual gas molecules\footnote{The sampled frequency is an average over many axial trajectories within the magnetic trap, so at an instant collision, the spectrogram shows an energy loss as a discontinuous jump in frequency.}, while the noise background appears as a random distribution of power spectral density. In the absence of pile-up and Doppler-shifted sidebands \cite{phenom_paper,classifier_paper}, the energy of the event is directly extracted from the reconstructed start frequency of the first track in time. 
It is for this type of CRES image to which the application of our new technique is developed.

\subsection{Project 8 Baseline Reconstruction}
\label{subsc:baseline_reconstruction}

In Project 8, reconstruction of CRES events from spectrograms has thus far relied on point-clustering based approaches such as extensions of DB-SCAN \cite{dbscan}.
A robust clustering technique dubbed the Sequential Track Finder \cite{christine_thesis,dan_thesis} has been developed to reconstruct both $^{83\text{m}}$Kr and tritium events and has been successfully used to build the first CRES tritium spectrum and extract a neutrino mass upper limit \cite{t2_prl}. 
In the following we briefly summarize the baseline track and event reconstruction strategy leaving additional details to \cite{t2_prl,t2_prc}.

The first step in reconstruction is to scale and normalize the power deposited in each frequency-time bin by subtracting the average noise amplitude and dividing the result by the variance of the noise in that frequency slice over all time bins.
The outcome is a frequency-independent intensity measure (termed the normalized power spectral density) which is used as a filter, allowing only those above a configured minimum threshold to remain.
The resulting filtered spectrogram is scanned over increasing time slices and track objects are created as lines for all groupings of bins that meet proximity-based configured constraints.
Track candidates are finally either kept or discarded if they respectively obey or fail further frequency- and time-gap tolerances set by the user.

Every surviving track has its start and/or endpoint sequentially discarded until the respective bin SNR exceeds a minimum ``trimming'' threshold; for this analysis, the threshold was pre-configured to SNR $= 6$. 
To fix actual tracks which have been erroneously identified by the algorithm into multiple sub-tracks, a clustering technique is further employed at the track-level to merge two or more of them only if certain mutual conditions are met e.g. similarity in track slope or overlaps frequency and time. 
Finally, a straight line is fitted to each remaining candidate in the spectrogram and multiple of these are grouped into an event following a head-to-tail matching in time between frequency scatters. 
The resulting first track in time designated as the start of the event. 

In order to control the false positive reconstruction rate, a further cut on normalized power spectral density is applied at the event level: a minimum threshold is set to dynamically vary depending on the number of tracks in the event and the number of bins in the event’s first track. 
Threshold values are lessened when the first track consists of more bins and is followed by many scattered tracks, effectively giving higher true positive confidence to a reconstructed event if there is evidence that it is more spatially dense (both locally and globally). 
As will be described later, control over the false positive event rate via these thresholds allows us to tune the baseline algorithm and perform a direct comparison to the ML model. 

In total, the baseline reconstruction consists of 29 configurable parameters whose values are manually set and optimally inferred from analysis of: the distributions of average power in events' first tracks, the total number of false events reconstructed from noise-only data, and a ``target'' false event rate among others.

%%%%%%%%%%%%%%%%%%%% Chapter 3: ML Model for Reconstruction %%%%%%%%%%%%%%%%%%%%
\section{Machine Learning Model for Event Reconstruction}
\label{sc:ML}

\begin{figure}[ht!]
\begin{center}
\includegraphics[width=0.8\textwidth]{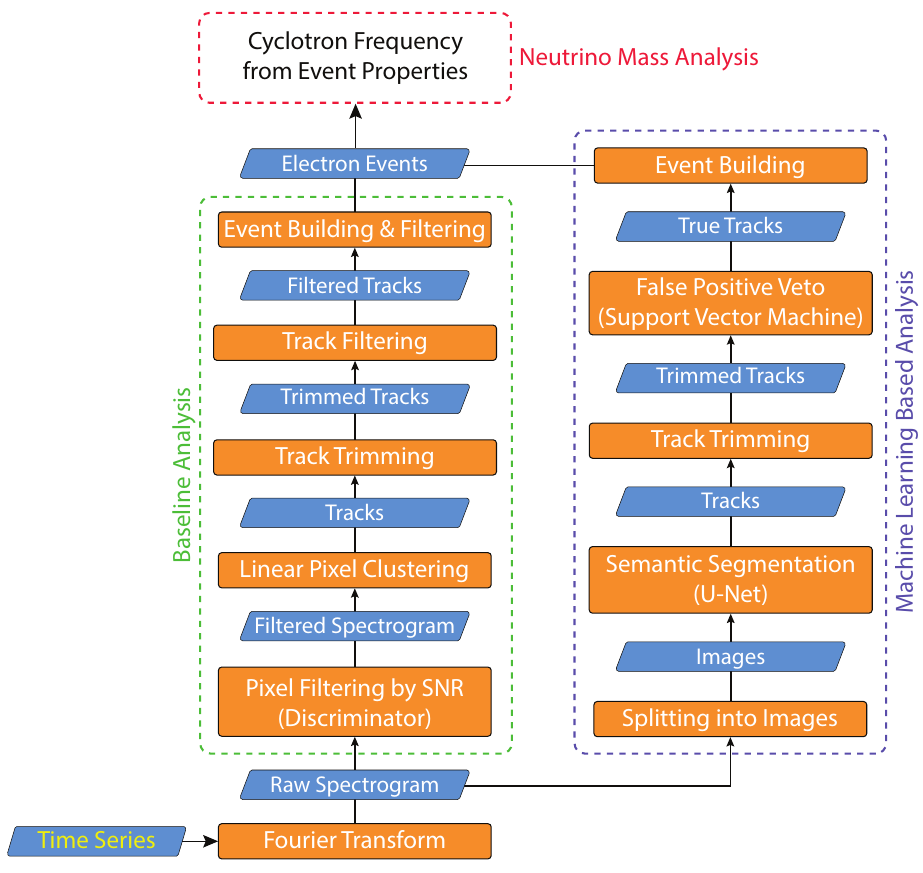}
\caption{A modular illustration of the proposed ML-based model for CRES event reconstruction (right dashed blue box) versus the existing baseline approach (left dashed green box) as applied in the Project 8 experiment. \label{fig:ml_model}}
\end{center}
\end{figure}

In this section we describe a new event reconstruction model whose backbone is the application of a CNN for track segmentation and an SVM which serves as a false positive veto. 
The motivation to employ a CNN-based deep learning approach described in Sec.~\ref{sc:intro}, plus the demonstrated achievements such class of models offer including highly-accurate localization of foreground images, robustness against noise, and relative ease of optimization given commercially available hardware \cite{MO2022}, make the ML technique a strong potential candidate for CRES signal reconstruction.
An illustration of the complete proposed model chain is shown in Figure \ref{fig:ml_model} and its submodules described in the following text. Discussion of the generation and properties of the simulated data to be used for training and validation will be left to Section~\ref{sc:simulation}.

\subsection{Spectrogram Segmentation}

\begin{figure}[ht!]
\begin{center}
\includegraphics[width=0.8\textwidth]{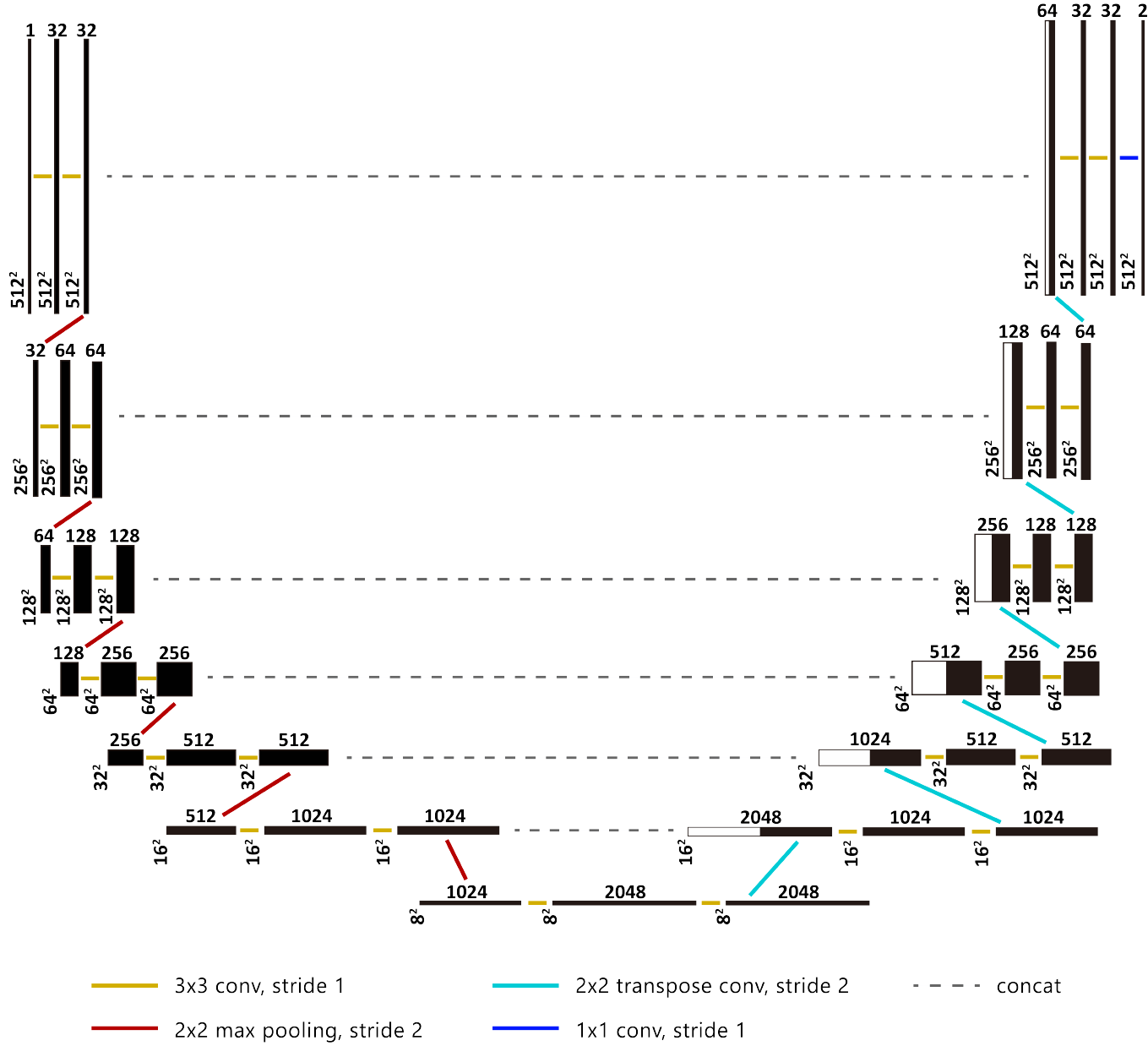}
\caption{An illustration of the deeper U-Net variant used for CRES signal segmentation. Blocks in black represent input layers (either images or feature maps) while those in white their respective concatenated versions. Input spatial dimensions are specified by vertically aligned integers (squared) while layer dimensions by horizontally aligned integers. The group of operations on the left side of the ``U'' defines the encoder while the group on the right defines the decoder. \label{fig:unet}}
\end{center}
\end{figure}

A typical CRES event in the Project 8 Phase II experimental setup roughly occupies a frequency span of 24 MHz and is fully contained within a 21 ms window of time. 
In this space, pixels belonging to track signals constitute less than 0.01\% of the spectrogram image on average\footnote{This ratio could be even lower depending on the size of the acquisition window, which could span a space significantly larger than a single CRES event.}, revealing a strong class-imbalanced classification problem with respect to the noise-filled background. 
To accurately segment these data-sparse images, we employ a deeper variant of the CNN U-Net architecture, first proposed in 2015 for biomedical imaging of neuronal structures in stacks of electron microscopy images \cite{unet}; see our model schematic in Figure \ref{fig:unet}. 
The U-Net architecture has been previously shown to produce very accurate segmentation with detailed spatial resolution in class-imbalanced problems \cite{isbi_challenge,channel_unet,3d_unet,unet_id}.
The key to success in achieving high spatial resolution is the novel sharing of ``skip connections'' from its \emph{encoder} side (convolutional feature extraction) to its \emph{decoder} side (spatial amplification with transpose convolutions).

In our application, the U-Net inputs are unpadded single-channel simulated CRES images of size 512 by 512 pixels (or 12.5 MHz by 21 ms) where the intensity of a pixel is the normalized power spectral density deposited in each frequency-time bin. 
Each input is accompanied by an equally-sized single-channel array of ground truth labels where 0 labels background and 1 labels track pixels.
The pixel intensities are transformed to follow a standard normal distribution (standard scaled).
It is important to note that the utilization of a CNN for track reconstruction advantageously offers an architecture that remains unaffected by the geometrical form of the input CRES signal (spectrograms with single or multiple-events, semi-linear or even curved tracks\footnote{CRES tracks featuring a prominent degree of geometric curvature may appear when charged particles emit cyclotron radiation at a resonant frequency of the confinement apparatus (such as in a cavity). This results in a significant loss of power over a prolonged period of time which changes the frequency of the signal continuously over time within a small window.}) or the dimensionality of the input image (2- or 3-dimensional).
The only model parameters which would require any alteration are the values of internal operations which are customized to produce a desired output shape and size.
This is an advantage not present in the baseline approach whose reconstruction steps and parameters assume a very specific signal form. 

In the U-Net architecture, the encoding arm makes use of the same basic operation many times over:
\begin{equation}
    \begin{aligned}
    \text{Encoder Unit}(N\text{ filters}) &\equiv \overbrace{3\times 3\text{, stride }1}^{\text{Convolution}\times N} \blacktriangleright \overbrace{3\times 3\text{, stride }1}^{\text{Convolution}\times N} \blacktriangleright \overbrace{2\times2\text{, stride 2}}^\text{Max-Pooling},\text{ where}\\
    \blacktriangleright &\equiv \text{ReLU} \rightarrow \text{Batch Norm.} \rightarrow \text{Dropout.}
    \end{aligned}
    \label{eq:encoder_unit}
\end{equation}
Similarly, the decoding arm uses:
\begin{equation}
    \begin{aligned}
        \text{Decoder Unit}(N\text{ filters}) \equiv \overbrace{2\times 2\text{, stride }2}^{\text{Transpose Conv.}\times N} \blacktriangleright \text{Concat.} \blacktriangleright \overbrace{3\times 3\text{, stride }1}^{\text{Convolution}\times N} \blacktriangleright \overbrace{3\times 3\text{, stride }1.}^{\text{Convolution}\times N}
    \end{aligned}
    \label{eq:decoder_unit}
\end{equation}
Readers unfamiliar with ML operations may refer to textbooks such as \cite{deep-learning-Goodfellow-et-al-2016} for terminology. 
In our implementation, both arms meet after the application of six encoder units at a spatial resolution of 8 by 8 pixels with 1024 feature maps. 
This ``middle'' portion furthermore consists of two more convolutions of the same type as Eq.~\eqref{eq:encoder_unit} without the max-pooling step. 
All filter weights are He initialized \cite{he_init} in order to constrain the hidden-layer variance to unity following the ReLU activation. 
While CRES tracks are essentially ``featureless'' when compared to neuronal structures, they are also confined to a very small area of the entire image. 
Therefore, we use additional encoding units to achieve a higher receptive field of our small-range pixel structures at the expense of extracting some number of redundant filters. 
The depth of the network (number of layers) will be treated as an architectural hyperparameter in Section~\ref{subsc:ml_optimization}.

During training, we found it necessary to apply batch normalization after the ReLU activation in $\blacktriangleright$ (as in Eq.~\eqref{eq:encoder_unit} and Eq.~\eqref{eq:decoder_unit}) in order for the losses to suitably converge. 
The addition of dropout layers allowed us to introduce regularization which improved generalization and avoided overfitting; optimization results will be discussed in Section~\ref{subsc:ml_optimization}. 
In the final steps of this module, feature maps are put through a convolution with two segmenting filters: one for track and one for background pixels, making the prediction of the network a 2-channel image of size 512 by 512 pixels with logits as intensities. 
To produce the desired segmentation mask, we apply a softmax function followed by an argmax function that selects one class per pixel following the highest probability. 
The resulting 1-channel mask may be directly compared to the respective ground truth mask for accuracy. 

\subsection{Building a Loss Function for Segmentation}
\label{subsc:building_loss}

Within confinement in the waveguide and magnetic trap, CRES electrons kept at a nominal pressure of $1.6 - 2.6 \times 10^{−6}$ mbar often scatter off residual H$_2$ molecules, the leading source of background gas in the experiment; for further details on full experimental gas composition see \cite{t2_prc}. This process incurs a minimum electron energy loss of approximately 11 eV (0.60 MHz), resulting in a wide inter-track separation of 23 pixels. Because of the large gap between foreground objects, we discard the border definition and weighting scheme of the original architecture \cite{unet}. Instead, due to the large class imbalance, we use a pixel weight
\begin{equation}
    w(\mathbf{r}_i) = \frac{1}{\text{percentage of }\mathcal{C}(\mathbf{r}_i)\in\text{Image}}
    \label{eq:unet_weights}
\end{equation}
for the loss where $\mathcal{C}(\mathbf{r}_i)$ is the ground truth class label of pixel $i$ at position $\mathbf{r}$ in the image. The weight is added to the pixel-wise softmax cross entropy loss in order to suppress penalty bias from the statistically boosted background. 

A further class imbalance is present in CRES spectrograms which is due to varying track topology within events. 
Recall that a given spectrogram may contain an event consisting of a single track or one made up of multiple scattered tracks. 
Generally, longer tracks are easier to reconstruct than short ones due to the consistency of the narrow signal profile over time, regardless of SNR. 
Conversely, segmentation is more challenging for short tracks which may resemble clumps of randomly adjacent noise pixels in the other extreme. 
In light of this prominent feature variation within events/spectrograms, efficient training demands that the loss account for penalizing misclassification of hard-to-classify pixels more severely than easy-to-classify pixels. 
Thus, we introduce a focal modulation term \cite{focal_loss} to define a compound total pixel-loss over two classes per each image:

\begin{equation}
    \text{Loss } = -\sum_{\mathbf{r}_i\in\text{Image}} (1-p)^\gamma w(\mathbf{r}_i)\log(p) + p^\gamma w(\mathbf{r}_i)\log(1-p) 
\end{equation}
where $p$ is the class probability for track pixels (foreground), $w(\mathbf{r}_i)$ is defined in Eq.~\eqref{eq:unet_weights}, and the sum is over all pixels $i$ at location $\mathbf{r}$ in the image. The new hyperparameter $\gamma$ controls the shift in penalty due to the imbalance, with higher values suppressing the loss for confident predictions ($p\gg 1$) and boosting it for uncertain ones ($p\ll 1$).

\subsection{Track Instance Segmentation and Trimming}

We produce track objects from the predicted segmentation masks with the use of scikit-image \cite{scikit-image}. 
Specifically, we employ the morphological operations of the \texttt{measure} submodule to group connected pixel areas (nearest neighbor pixels in all directions) into track instances, obtain their coordinates, and fit a line to each using a least squares estimation. 
In order to account for over-coverage segmentation of foreground in the direction parallel to the track (from which the start time is drawn), we introduce the same trimming procedure and trimming threshold as the baseline approach for both ends (see Section \ref{subsc:baseline_reconstruction}). 
The result is a reduction of start time over-coverage from approximately 3.5 pixels to less than 0.1 pixel. The start frequency of each track is then calculated using the intersection of the fitted line with the start time bin.

\subsection{False Positive Veto and Event Reconstruction}
\label{subsc:fp_veto_event_rec}

A false positive is a model inference which erroneously labels and groups adjacent background pixels as a track. 
The mispredictions encountered largely encompass the short track regime where occasional patches of noise with relatively high SNR resemble true short tracks. 
Physically, the length of a track is constrained by the gas pressure within the confining waveguide: higher pressure leads to frequent particle scattering and, thus, shorter signal duration. 
The converse occurs for lower pressures, resulting in longer tracks. 
In Project 8 Phase II, the gas pressure was optimized to balance both the number of detected decay events and the scattering rate controlling the number of electron-residual gas collisions \cite{t2_prc}. 
Under this scheme, the shortest tracks allowed could be as small as 3 pixels in length. 
In our simulation of training images (described in Section \ref{sc:simulation}), we also allowed for events with equally short tracks. 
In light of this inclusion, the number of false positive tracks predicted by the semantic + instance segmentation modules averaged to about 1 false track every 13 images.

To lower the false track prediction rate, we introduce a radial-basis kernel SVM trained to serve as a false positive veto. The input to the SVM are single track objects represented by a 4-dimensional vector composed of: track slope, track SNR, track length, and start time. In general, the use of SNR and track length as discriminating features greatly improves the classification accuracy for long tracks with sharp profiles. The slope also acts as a powerful discriminatory feature for those segmented tracks which do not follow physical constraints: a true CRES track must have a positive slope while a false track may be flat, vertical, or have a negative incline. Additional use of the start time allows for removal of false positives found near or at the vertical edges of the image with all such regions strictly excluded during simulation. The SVM veto module is able to greatly reduce the false positive track rate to 1 false track every 233 images\footnote{Further improvement in false positive rate is seen with the addition of the event-builder module which completes the reconstruction sequence (see Table \ref{tb:efficiency_comparison}).}. Full events are then reconstructed from remaining tracks using the same head-to-tail matching constraints as the baseline approach described in Section \ref{subsc:baseline_reconstruction}. The first track in time is taken as the start of the event.

%%%%%%%%%%%%%%%%%%%% Chapter 4: Simulation of CRES Events %%%%%%%%%%%%%%%%%%%%
\section{Simulation of CRES Events for Optimization}
\label{sc:simulation}

A significant advantage of using the simulated data described below for training and testing a model is the confident knowledge of ground truth properties. To avoid simulating CRES events with traditional particle-tracking calculations which demand great computational power and processing times, we developed a simpler and more efficient parametric-based Monte Carlo approach within the Locust software package \cite{locust_paper}. The Locust software models the response of an antenna and receiver chain to time-varying electromagnetic fields with the use of internal classes called ``generators''. For our application, the newly developed \texttt{LMCFakeTrackSignalGenerator} produces a CRES event of desired structure and duration by sampling individual track parameters from a number of different probability density functions (PDFs) at run time. For example, lengths of tracks within an event are drawn as samples from an exponential distribution with configured mean. A thermal noise floor may also be added to the simulation by drawing random voltages from a normal distribution of configured mean power. The combined signal plus background are processed through a simulated Project 8-like antenna module, RF receiver, and data acquisition chain to produce a raw time series and then a spectrogram. Tens of thousands of simulated spectrograms may be produced within just a few CPU hours.

\subsection{Validation of Simulated Data}
\label{sc:sim_validation}

To obtain a physically realistic SNR distribution from which we simulate signal intensity, we first perform a one-shot particle-tracking simulation with Kassiopeia \cite{kassiopeia_paper} of 17.8 keV $^{83\text{m}}$Kr electrons in a harmonic magnetic trap of 1.4 mT depth, a relatively deep magnetic field trap configuration which increases effective volume for electron trapping \cite{t2_prl}. Electrons are given random starting positions along detector boundaries and pitch angles of $\theta\geq \ang{89}$ to remain consistent with trapping limitations\footnote{The pitch angle is the angle between the electron's momentum and the magnetic field direction at the bottom of the trap and directly constraints magnetic trapping.}. Particle-tracking occurs over a duration of \SI{40.96}{\micro\second} (the exact length of one spectrogram time bin) from which the incident electromagnetic fields are mixed and sampled using Locust. A fixed gain scales the resulting power, and the SNR distribution with approximate mean SNR $=3$ is finally retrieved relative to the noise floor. Note that this SNR distribution is only used to configure the power of first tracks in Locust-simulated events. 

For subsequent tracks, \texttt{LMCFakeTrackSignalGenerator} internally draws pitch angles from the electron-H$_2$ inelastic scattering differential distribution of Rudd \cite{rudd} and the scattered track power is calculated using the power-pitch angle relation given by the phenomenological CRES model \cite{phenom_paper}. As the condition for trapping relies on a minimum pitch angle value, the scattering distribution also implicitly controls the total number of tracks per event through the scattering scale parameter $G_3$, a fitting parameter of the cross section model \cite{rudd} to be inferred from the data. The size of the frequency jump between scatters is dictated by losses in energy taken from oscillator strength data \cite{lxcat} together with a Bethe inelastic-scattering theoretical model for low energies (< 50 eV). Extrapolation to arbitrarily high energy losses (> 50 eV) is done using the model from Aseev \cite{aseev_energy_loss}. 

We tune the remaining simulation parameters to match realistic events by studying the track and event properties of 6300 real Phase II 17.8 keV $^{83\text{m}}$Kr electrons events born within a 1.4 mT-deep quadrupole-harmonic trap~\cite{christine_thesis}. Using the standard baseline method, reconstructed track properties such as slope and length were fit to their respective statistical distributions and the relevant moments were extracted for use in simulation (see Table \ref{tb:locust_sim}). One important point to note is that the resulting reconstructed mean number of tracks per event is not expected to reflect the true underlying physical value since very short/low-SNR tracks are often missed during reconstruction. To find the true underlying number, multiple sets of tracks were simulated while varying the scattering scale parameter $G_3$. The number of reconstructed tracks per event was then compared to the reference, real, data until the value of $G_3=0.0064$ was found to be optimal in matching both. With this, the simulated average number of tracks per event is 5.1.

\begin{table}[H]
\caption{\label{tb:locust_sim} Simulated event properties' constraints including source PDFs where applicable. Values were chosen in order to simulate data which matches real Project 8 Phase II CRES events.}
\begin{indented}
\lineup
\item[]\begin{tabular}{@{}*{3}{l}}
\br
\0\0Parameter & PDF & Value(s)\cr
\mr
Start Frequency & Uniform & Between 25.9060 and 25.9065 GHz\cr
First Track Power & Exponential & Drawn from Kassiopeia simulation \cr
%Start Pitch Angle & Fixed & $90^{\circ}$ \cr
Scattering Pitch Angle & $(1+\cos^2(\diff\theta)/G_3^2)^{-1}$ \cite{rudd} & $G_3 = 0.0064$\cr
Start Time & Uniform & Between \SI{50}{\micro\second} and \SI{13}{\milli\second}\cr
Track Length & Exponential & $\lambda^{-1} =$ \SI{0.18}{\milli\second}\cr
\multirow{2}{*}{Track Slope} & \multirow{2}{*}{Normal} & $\mu = 0.3523$ MHz/ms \cr 
& \, & $\sigma = 0.0545$ MHz/ms \cr
%Number of Tracks per Event & Geometric & $p^{-1}=5.1$ tracks\cr
No. of Events per Simulation & Fixed & 1 \cr
Magnetic Field & Fixed & $B = 0.9578$ T \cr
\multirow{2}{*}{Thermal Noise Power Spectral Density} & \multirow{2}{*}{Normal} & $\mu = 3.0\times 10^{-14}$ W/Hz \cr 
& \, & $\sigma = 1$ W/Hz\cr
\br
\end{tabular}
\end{indented}
\end{table}

\begin{figure}[ht!]
\begin{center}
\includegraphics[height=5.25cm]{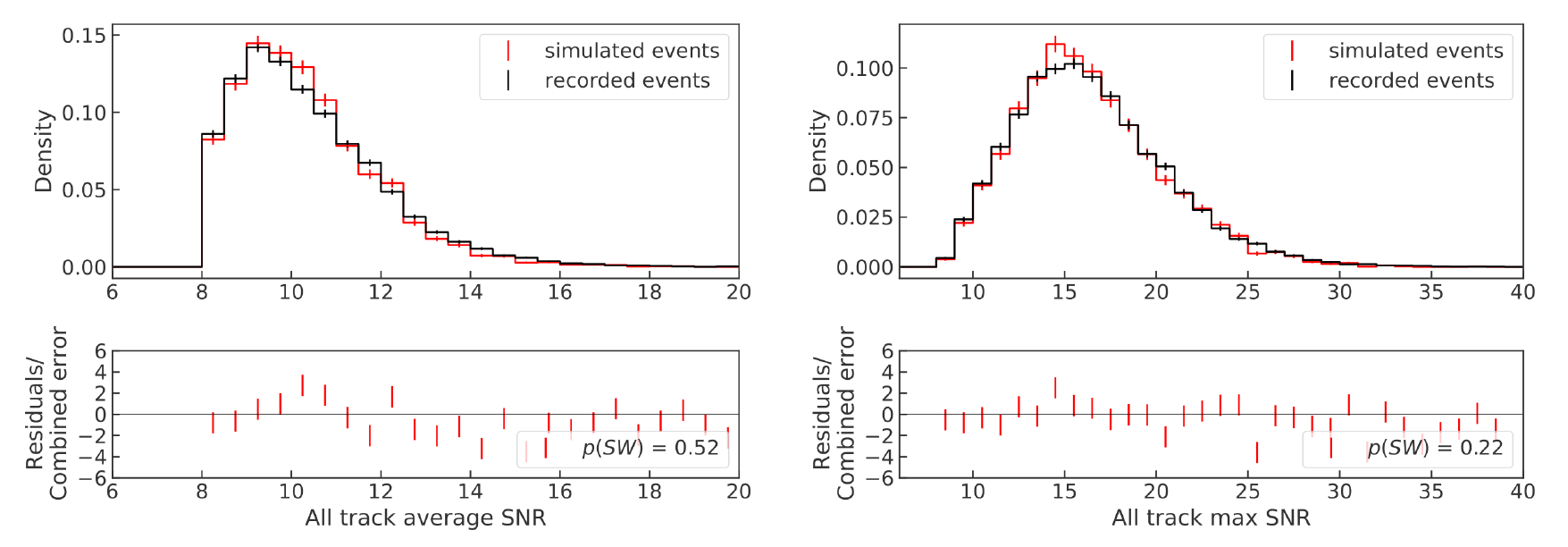}
\caption{Comparison of average SNR (left) and max SNR (right) between simulated and real tracks showing the equivalence between simulation and reality \cite{christine_thesis}. Accompanying p-values for the Shapiro-Wilkes test on residuals, $p(SW)$, are shown in the lower legends. \label{fig:sim_equivalence1}}
\end{center}
\end{figure}

Equivalence between simulated and real events is shown by comparing a set of 92,400 generated fake events to 6,257 real events taken from 17.8 keV $^{83\text{m}}$Kr electron data. Most relevant to the use of spatially small convolutional filters in the ML-based approach is the validation of realistic SNR fluctuations along a simulated track. In Figure \ref{fig:sim_equivalence1} we show a comparison of the average SNR and maximum SNR of all tracks in reconstructed real and simulated events. Applying the Shapiro-Wilkes test \cite{SHAPIRO1965} to the residuals, we cannot rule out the equivalence between simulated and real data at 95\% significance ($p\gg 0.05$, see figure legend). 

\iffalse
We also investigate the SNR fluctuation on a detailed bin-by-bin basis and display the comparison of SNR standard deviation spread sorted by average track SNR in Figure \ref{fig:sim_equivalence2}. The agreement between simulated and real data suggests that the dominant source for inter-bin power fluctuation is the angle and relative position at which tracks cross frequency bins in combination with the fluctuating noise power added. From this, we conclude that the agreement between simulated and real data is appropriate, and confidently use simulation for training and testing the reconstruction models.

\begin{figure}[ht!]
\begin{center}
\includegraphics[width=0.9\textwidth]{sim_comparison2.pdf}
\caption{Comparison of track bin SNR standard deviation between real tracks (left) and simulated tracks (right) versus average track SNR \cite{christine_thesis}. \label{fig:sim_equivalence2}}
\end{center}
\end{figure}
\fi

\subsection{Creation and Selection of Training and Test Data}

With the simulated module vetted, we create a training set for model optimization by running two concurrent Locust simulations per event: one including noise and one without any noise. Besides the absence of noise in the latter, the two simulated events remain identical as long as the same random seed is configured at run time. Spectrograms are produced from Locust voltage time series using the Katydid software \cite{katydid} with each spectrogram measuring 4096 by 512 pixels (100 MHz by \SI{21}{\milli\second}) with physical dimensions of 24.41 kHz by \SI{40.96}{\micro\second} per pixel. As the U-Net architecture expects square inputs\footnote{Arbitrarily large inputs are allowed at the cost of additional computational power for which tiling or clipping strategies are usually employed \cite{unet,unet_id}.}, we vertically split each spectrogram into eight $512\times 512$ pixel square sections. Although the vast majority of events are confined to a single image, we keep track of the ordered image index (0-7) in order to correctly identify the true start of the event after classification. We are ensured that the first track in the event is not split in two by confining the simulated event start frequency to a range of $\sim 0.5$ MHZ (or $\sim 20$ pixels) within the first square\footnote{This effect does not alter the efficiency calculation in Section \ref{subsc:efficiencies} but may have an effect on the track property errors. As will be seen in the results of Section \ref{subsc:abs_track_errors}, no significant bias was detected.}. Images without at least a single track are discarded for segmentation training purposes.

The noiseless versions of simulated events are used only to create segmentation ground truth labels by assigning a value of 0 (background pixel) or 1 (track pixel) to those which fall below or are above a minimum intensity threshold respectively; a limit of 10\% the maximum pixel intensity in the image was found to be suitable for labeling. However, we found that allowing the ML model to train on pixels from very short tracks of equally low SNR introduced a strong bias that favored acceptance of false positives and false negatives. To account for this, we only select pixels from tracks with overall relatively strong signal profiles for training. Considering that the average bin noise power spectral density follows a gamma distribution for $N$ bins such that $X_\text{noise}\sim \Gamma(x;N,1/N)$ for random variable $x$, in order to see less than one false track in the entire set of training images we must have $\text{SF}(\Gamma(x;N,1/N))<1/N_\text{tot}$ where SF is the survival function and $N_\text{tot} = (\text{Locust event-simulation time})\times(\text{Number of frequency bins})\times(\text{Number of training images})$. This implies that $x > \text{SF}^{-1}(\Gamma(1/N_\text{tot},N,1/N))$, giving us a useful SNR vs. track length relationship to use as a minimum dynamic threshold when selecting training points.

\begin{figure}[ht!]
\begin{center}
\hspace*{-1.45cm}
\includegraphics[width=1.15\linewidth]{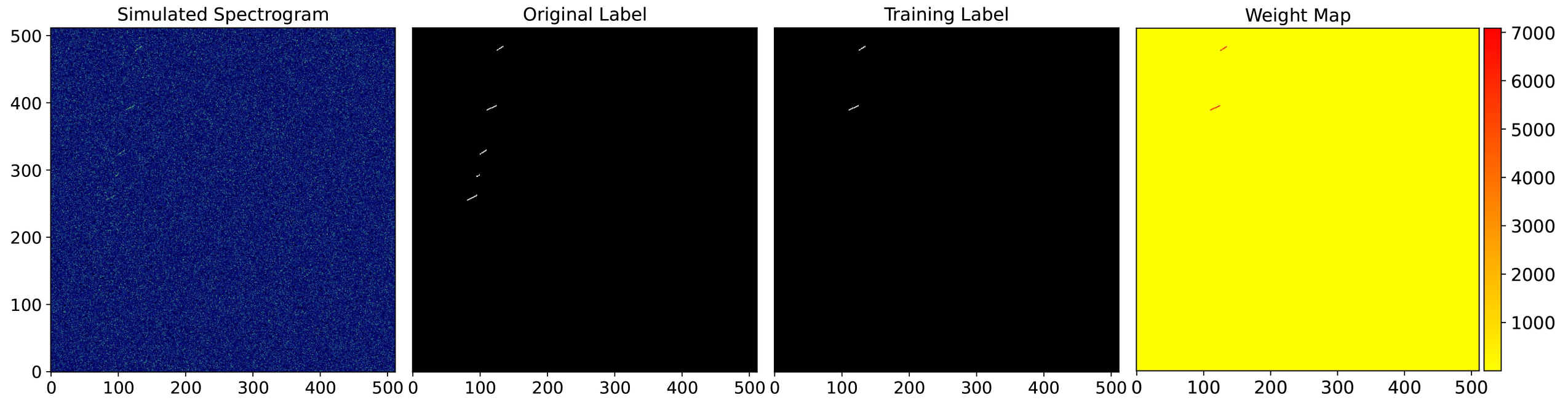}
\caption{An example of a training image (simulated spectrogram) including its original mask labels, training label, and weight map. The original labels serve as ground-truth for comparisons with predicted segmentations and track objects while the training labels and weight maps are used only during model optimization. The scales of x- and y-axes are (in) pixels. \label{fig:ex_tr_ims}}
\end{center}
\end{figure}

With the above parameters in place, we finally simulate 40,194 training images (51,401 tracks), 10,413 validation images (13,196 tracks), and 10,575 test images (13,522 tracks). This constitutes an approximate 80\%/20\% split between train and validation (together) and test sets. An example of a training image is shown in Figure \ref{fig:ex_tr_ims} where in the ground truth mask white represents track pixels and black represents background pixels. For the pixel weight map, only the relative values between classes are important and not the absolute scale (see Eq.~\eqref{eq:unet_weights}). The spectrograms, masks, and maps are stored as 2-dimensional arrays in HDF5 files for ease of interfacing with U-Net code. The pre-training-thresholding ``original label'' pixel masks are kept in order to compare the predicted track objects (not segmentation) to the simulated ground-truth.  

%%%%%%%%%%%%%%%%%%%% Chapter 5: Model Optimization %%%%%%%%%%%%%%%%%%%%

\section{Optimization of Reconstruction Models}
\label{sc:optimization}

In this section we describe training of the ML model and the strategy to tune the baseline model for comparison.

\subsection{Machine Learning Optimization}
\label{subsc:ml_optimization}

The implementation of the architecture in Figure \ref{fig:unet} was written in Python with the Tensorflow library \cite{tensorflow2015-whitepaper} using AdaDelta as the optimizer. All optimization was conducted on two NVIDIA Tesla P100 GPUs running in parallel and done separately for the U-Net and SVM. Optimal hyperparameters for the U-Net were found via a randomized grid-search using MLflow \cite{mlflow} with the following strategy: define an instance of the for each tuple of sampled hyperparameters, optimize over the entire training set, and quantify its success over the validation set using the pixel-wise $F_1$ score. Starting with 50 U-Net instances, the number was halved after every epoch by keeping only those with highest $F_1$ scores. The final, optimal, hyperparameters are listed in Table \ref{tb:hyperparam}.  

\begin{table}[h]
\caption{\label{tb:hyperparam} Optimal hyperparameters for U-Net and SVM modules including range of values explored and/or statistical moments defining the sampling distributions.}
\begin{indented}
\lineup
\item[]\begin{tabular}{@{}*{4}{l}}
\br
\0\0Hyperparameter & Module & Range Explored/Stat. Moment & Optimal Value\cr
\mr
Kernel Size & U-Net & $\{3,5,7\}$ & 3 \cr
$\gamma$ (modulation factor) & U-Net & $\{1,2,3\}$ & 2 \cr
Dropout Rate & U-Net & $\{0.0,0.1,0.2\}$ & 0.0 \cr
Network Depth & U-Net & $\{4,5,6,7\}$ & 6 \cr
$C$ & SVM & Exponential, $\lambda=0.01$ & 130.8 \cr
$\gamma$ (radial-basis) & SVM & Exponential, $\lambda=10$ & 0.054 \cr
\br
\end{tabular}
\end{indented}
\end{table}

In the average simulated track regime (0.18 ms in length and 0.35 MHz/ms in slope), a track crosses a width of about 3 pixels in height by 5 pixels in length. Here, the small kernel size (3) and large number of encoder units (6) use low-resolution feature maps to find the smallest possible signals which would otherwise be overtly- or multiply-contained by maps with much bigger receptive fields. A dropout rate of zero makes sense given the high class-imbalance as variance amplification is not expected to improve accuracy when there is $>99\%$ background-pixel abundance. Perhaps more telling of the nature of CRES events, a modulation factor of $\gamma=2$ tells us that a steep penalty-shift strategy is a necessity in order to overcome the large differences in track-composition between different events.

\begin{figure}[ht!]
\begin{center}
\includegraphics[width=0.9\textwidth]{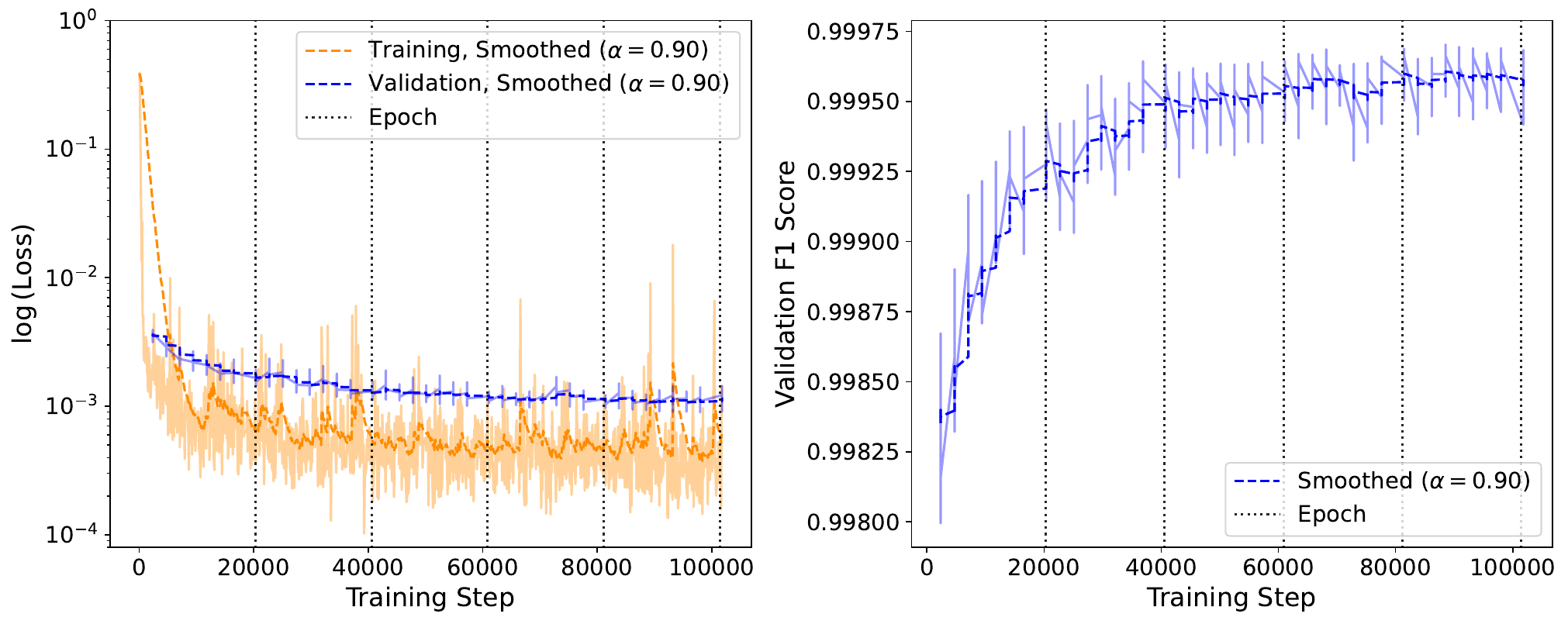}
\caption{Development of the training (orange) and validation (blue) losses (left) and validation $F_1$ score (right) over the optimization procedure. An exponentially weighted moving average (smoothed with $\alpha=0.9)$ is also displayed. \label{fig:unet_loss_f1}}
\end{center}
\end{figure}

With the optimal hyperparameters fixed, model training was performed with a mini-batch of two images and took five epochs to complete utilizing approximately 79 hours of wall-time with 45.06 GB of memory. The development of the training and validation losses, as well as the validation $F_1$ score can be seen in Figure \ref{fig:unet_loss_f1} where we employ an exponentially-moving average to track the overall behavior in light of the small batch size. In the figure, we see rapid progress in loss optimization and $F_1$ score during the first epoch, with subsequent epochs only featuring fractional gains. This is indicative of the ease with which the U-Net learns to classify background pixels as a result of their abundance, while very slowly succeeding in classifying the few track pixels present in the images. As a measure of robustness, we build a ROC curve and extract the area under the curve (AUC) as well. The final validation metrics after optimization are: $F_1$ score $= 0.9996$ and AUC $=0.9996$, representing a model with excellent classification accuracy and stability. For the test set of 10,575 images, segmentation took approximately 38 hours, i.e less than 0.22 seconds per image. Mirroring the strong validation result, the test $F_1$ score was 0.9986.

The SVM false track veto module was implemented in Python with the scikit-learn library \cite{scikit-learn}, trained with 9,468 tracks, and tested using 4,059 tracks. The SVM hyperparameters $C$ and $\gamma$ were obtained using a 2-fold cross-validation strategy with a randomized grid search and optimally found to be $C=130.8$ and $\gamma=0.054$ with $F_1$ score $= 0.9689$, see Table \ref{tb:hyperparam}. At testing, the final SVM model reached an $F_1$ score of 0.9680 and an AUC of 0.9868. The hyperparameters values represent a model which features a balanced variance ($C \gg 1$) and smooth classification boundary ($\gamma\ll 1$). 

\subsection{Baseline Tuning}
\label{subsc:optimization_baseline}

Since the ML-based method is not yet able to reach the zero false event rate reported in the Project 8 Phase II neutrino mass analysis \cite{t2_prl}, we tune the baseline parameters so that its false positive event rate matches that resulting from the ML-method. As shown in Section \ref{subsc:efficiencies}, this allows us to measure efficiency by comparing the accompanying true positive event rates. 

All baseline parameter configurations related to pixel clustering and track identification were fixed to be the same as in Phase II track and event reconstruction \cite{p8_neutrino2022}. For the remaining configuration, the minimum power thresholds for different number of tracks and first-track number of bins need to be determined so that the number of false positive events is adjusted to the desired value. For this, false positives events are divided in groups based on the number of pixels in the first track and the total number of tracks per event. First, test data was processed with the ML method to find the number of false events reconstructed. Then, we manually tuned the the baseline power thresholds per group until we achieved the same false event in the same set of data.

Because there was just one false event detected using each method (see next section), the false event rates shown here are only approximately equivalent. Since this study serves as a proof of concept comparison between the ML and baseline models, we did not pursue more intensive tuning of the baseline method that would have allowed for a more reobust comparison.

%%%%%%%%%%%%%%%%%%%% Chapter 6: Results and Comparison %%%%%%%%%%%%%%%%%%%%

\section{Results and Comparison to Baseline Approach}
\label{sc:results}

\begin{figure}[H]
\begin{center}
\includegraphics[width=\textwidth]{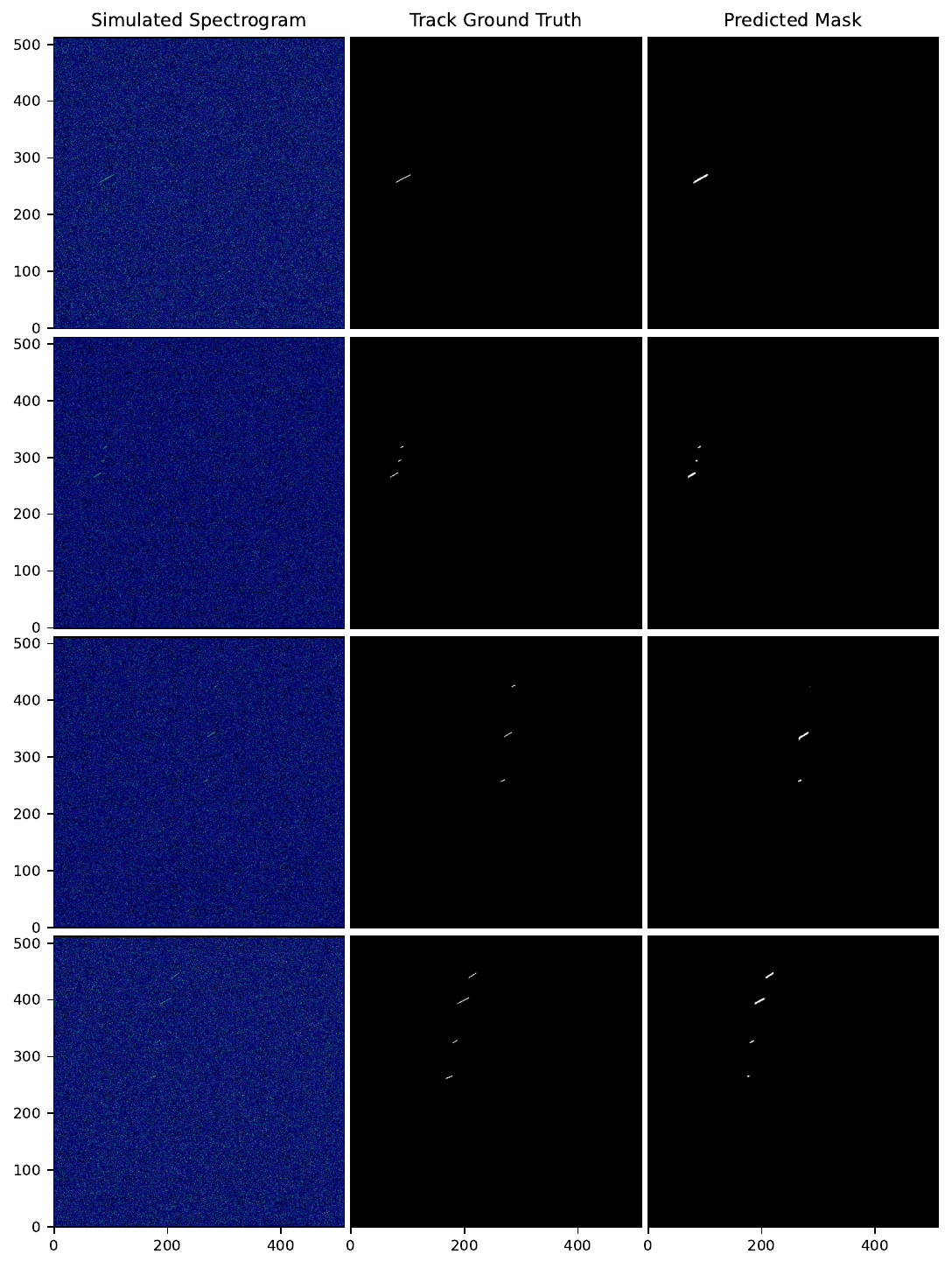}
\caption{Four simulated CRES events (one per row) alongside their ML model reconstruction. Reconstruction is performed on single and multi track events (first and second examples) with precise localization. In the third example, the true event start is correctly reconstructed although the last track in the event is missed. The last example shows the reconstruction of an event start (first track) that is shorter than ground truth. In general, all reconstructed tracks are slightly wider than ground truth. The scales of x- and y-axes are (in) pixels. \label{fig:ml_events}}
\end{center}
\end{figure}

Since the event building procedure is equivalent in both reconstruction approaches, we first investigate the results of the track reconstruction modules on the test set of images, comparing both to the ground truth and to each other. Then, we move on to compare event reconstruction results on the same test set and focus on efficiencies as a final measure of performance. As a visual reference for the ML model results, we display four reconstructed events from the test set in Figure \ref{fig:ml_events}.

\subsection{True and False Positive Tracks}

\begin{figure}[ht!]
\begin{center}
\includegraphics[width=0.5\textwidth]{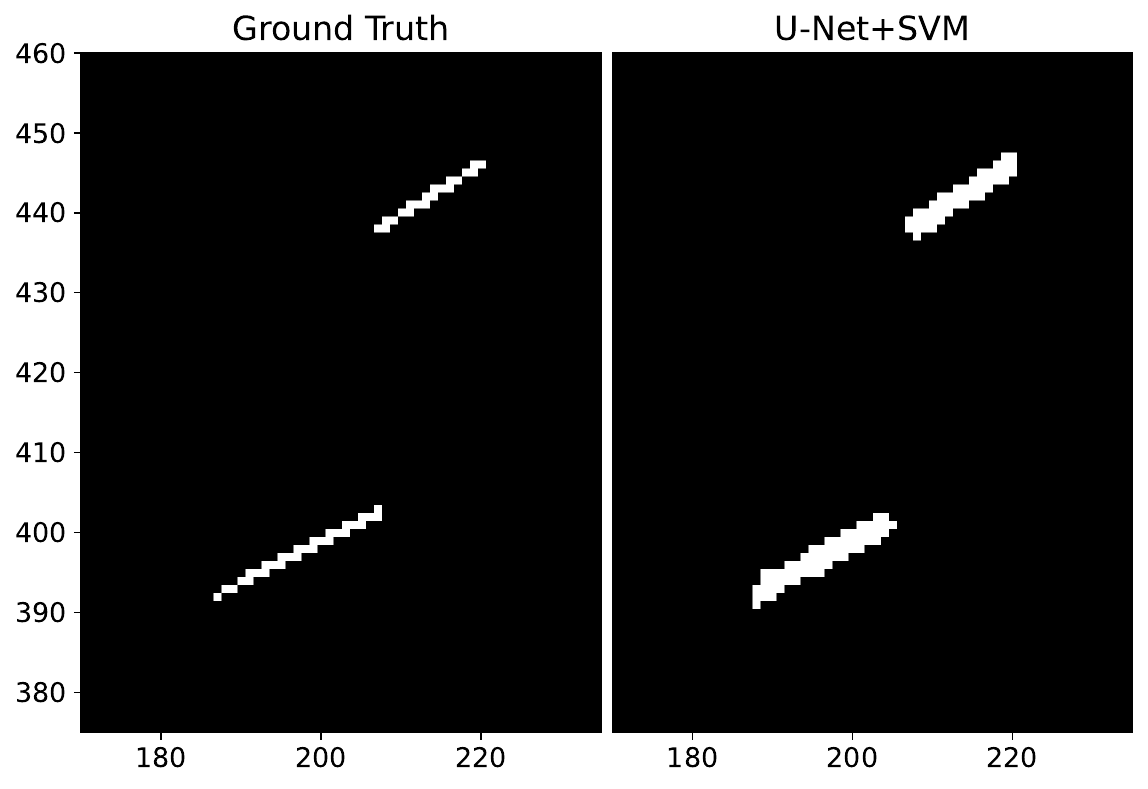}
\caption{An example comparison between a ML reconstructed track (right) and its respective ground truth mask (left). Most ML track predictions tend to over-cover the truth mask in the direction normal to the track, resulting in ``wider'' tracks. On the other hand, the trimming procedure tends to over-cut tracks in the parallel direction, making them shorter; see Section \ref{subsc:abs_track_errors} for quantitiative comparison. The scales of x- and y-axes are (in) pixels. \label{fig:ml_track_zoom}}
\end{center}
\end{figure}

Since the track object prediction from the ML model is a region of connected pixels, we define a true positive as a track whose pixel area has a Jaccard index \cite{Jaccard1912} greater than 0 with respect to the same pixel area in the ground truth mask. Conversely, a false positive is a track whose Jaccard index is exactly zero. However, in order to discern morphological discrepancy between under- and over-coverage cases, we introduce the mismatch index between images (or sets of pixels) $A$ and $B$:
\begin{equation}
    M(A,B) \equiv \begin{cases}
             1-|A\cap B|/|A| \quad &\text{if }B\text{ under-covers }A \\
             0 &\text{if }A\text{ and }B\text{ perfectly match} \\
             |A\cap B|/|B| &\text{if }B\text{ over-covers }A.
    \end{cases}
\end{equation}
with range $(-1,1)$, giving under-coverage a negative score, a perfect match a score of zero, and over-coverage a positive score. Within the set of true positives the majority (approximately $62\%$) have a positive mismatch index with average 0.49 and the rest a negative mismatch index with average -0.32. This tells us that most ML track predictions tend to be larger than the ground truth with further investigation revealing that the over-coverage is in the direction normal to the track slope. In general, predictions hold an average width about 1.5 times larger than the truth (see Figure \ref{fig:ml_track_zoom} as example). The under-coverage cases are almost all completely due to trimming the track ends (see next section), which uses the same pre-configured parameter as the baseline. In the future, the over-coverage seen here could possibly be improved by the inclusion of a Lagrange multiplier constraint for track width in a track-object based loss.

\newpage
\newpage

Tracks resulting from the baseline reconstruction are one-dimensional lines fitted to individual pixels. Thus, we define a rectangular bounding box with the reconstructed track as its diagonal and compare this region to the same area in the truth mask. If there is any overlap in the box region with a ground truth track (also with its own bounding box), the reconstructed track is labeled a true positive. A false positive results when there is no overlap between masks. Due to the same trimming procedure, the baseline tracks also tend to be shorter in the direction parallel to the track slope. 

\subsection{Absolute Track Property Errors}
\label{subsc:abs_track_errors}

We quantify the accuracy and precision of both reconstruction methods by inspecting the differences in reconstructed track property $P_i$ (e.g. track slope, start time) between true positive track prediction and ground truth with the directional error
\begin{equation}
    \epsilon_{P_i} = P_{i,\text{prediction}} - P_{i,\text{ground truth}},
\end{equation}
keeping the sign in order to capture over- or under-estimation. Table \ref{tb:tr_error_comparison} shows the mean error for all track properties of importance, alongside their sample standard deviation.

\begin{table}[h]
\caption{\label{tb:tr_error_comparison} Reconstructed mean track parameter errors $\pm$ one sample standard deviation, comparing the ML to baseline reconstruction results.}
\begin{indented}
\lineup
\item[]\begin{tabular}{@{}*{3}{l}}
\br
\0\0Track Parameter & U-Net+SVM Error & Baseline Error\cr
\mr
Start Time & \hspace{0.005cm} $0.003\pm0.082$ ms & \hspace{0.005cm} $0.057\pm0.093$ ms \cr
End Time & -$0.002\pm0.078$ ms & -$0.101\pm0.157$ ms \cr
Track Length & -$0.005\pm0.113$ ms & -$0.158\pm 0.179$ ms \cr
Track Slope & -$0.013\pm0.079$ MHz/ms & \hspace{0.005cm} $0.002\pm0.066$ MHz/ms \cr
Start Frequency & \hspace{0.005cm} $0.003\pm0.031$ MHz & \hspace{0.005cm} $0.020\pm0.034$ MHz \cr
End Frequency & -$0.002\pm0.027$ MHz & -$0.037\pm0.058$ MHz \cr
\br
\end{tabular}
\end{indented}
\end{table}

Recall that the track start and end times are directly defined by the first and last (horizontal) pixel bins, each bin with a physical-equivalent size of \SI{40.96}{\micro\second}. Therefore, to one standard deviation, the start time error from the ML reconstruction can be as much as 2 pixels, whereas the baseline error can vary up to 4 pixels. To the same precision, the ML end time error can be as much as 2 pixels while the baseline error can be as much as 6 pixels. Taking the sign of the errors into account, we conclude that both reconstructions produce tracks which start too late and end too early, resulting in a negative mean track length error. However, the baseline track length errors remain larger (up to about 8 pixels) than those of the ML method (up to about 3 pixels). 

For the track slope, Table \ref{tb:tr_error_comparison} shows that the baseline makes a better estimate. First, we can attribute this to over-coverage of tracks in the direction normal to their profile by the ML model, which results in a thicker track width. This latter degree of freedom is not present in the 1-dimensional reconstruction approach of the baseline method. Note that the start and end frequencies, which are directly determined from intersections of the slope with the respective start and end time bins, have errors which represent a mixture of both time and slope quantities. Second, the baseline analysis makes explicit use of the pixel SNR as a weight for slope regression while the ML approach does not; this adjustment minimizes the error by boosting the importance of true-positive-like pixels over true-negative-pixels in the fit. 

Although the mean slope error is significantly lower in the baseline method, the much smaller mean error on the start and end times in the ML-analysis are enough to produce more accurate estimates of the track frequencies. For start frequencies, the mean errors (alongside standard errors of the mean) are $\mu_\text{ML}=(0.003\pm 0.002)$ MHz and $\mu_\text{Baseline}=(0.020\pm 0.002)$ MHz, showing an almost order of magnitude improvement in accuracy. For precision, however, we look at the standard deviations (and their standard errors): $\sigma_\text{ML}=(0.031\pm0.001)$ MHz and $\sigma_\text{Baseline}=(0.034\pm0.001)$ MHz. Recalling that a frequency bin is about 24 kHz across, these results constitute an average offset of 1.3 (ML) to 1.4 pixels (baseline), indicating that both methods tend to reconstruct tracks further along the vertical in the image rather than earlier (i.e at higher frequencies). 

Ultimately, the mean start frequency error results in a magnetic field miscalibration which affects both $^{83\text{m}}$Kr and tritium data. This effect incurs an energy shift of approximately 0.06 eV (ML) to 0.42 eV (baseline) on the either spectrum, showing a favor towards higher accuracy in the former. On the other hand, the energy resolution (and therefore spectrum smearing and neutrino mass value) is impacted by the spread of the errors (i.e precision) which are $\sigma_\text{ML}=(0.655\pm0.021)$ eV and $\sigma_\text{Baseline}=(0.718\pm0.021)$ eV when converted to energy. From this, we conclude that both methods show comparable sub-eV resolution and stand as viable candidates for high-resolution CRES spectroscopy. 

\subsection{Efficiencies}
\label{subsc:efficiencies}

A straightforward measure of efficiency is to compute the fraction of all simulated events which were correctly reconstructed by each method, where the comparison to ground truth is made with first tracks in events. However, Doppler shifting of the cyclotron signal at the antenna results in a total radiated power which is shared between main carrier and sidebands of different frequencies \cite{phenom_paper}. The relative strength of the sideband signals depends on the pitch angle of the electron, which in turn is minimally bound by the shape of the magnetic trap \cite{phenom_paper}. If conditions are such that the pitch angle is significantly different from \SI{90}{\degree}, the main carrier power may be low enough to be undetectable. 

In Project 8 Phase II real data, reconstructed events were observed to feature SNR values varying between 4 and 20, consistent with events of pitch angles greater than \SI{89.33}{\degree}. In our simulated data, the track power was sampled only for pitch angles larger than \SI{89}{\degree}. Taking the ratio of the effective trapping volumes between minimum pitch angles \SI{89.33}{\degree} and \SI{89}{\degree}, we expect only about $44\%$ of all simulated events to be visible (detectable) after trapping conditions are satisfied. Furthermore, following theoretical modeling of CRES signals with the Viterbi algorithm \cite{viterbi_paper}, the minimal reconstructable track length given Phase II conditions is approximately 3 pixels, diminishing the percentage of visible electrons by $59\%$. Finally, making a cut at the minimal Phase II reconstructable SNR of 4 as a realistic approximation, we keep only $76\%$ of this fraction of events. Thus, we expect that in total only about 20\% out of all simulated events are actually reconstructable by either method.

\begin{table}[h]
\caption{\label{tb:efficiency_comparison} Statistics for reconstructed events including efficiency for ML-based and baseline methods.}
\begin{indented}
\lineup
\item[]\begin{tabular}{@{}*{3}{l}}
\br
\0\0 Quantity of Interest & U-Net+SVM & Baseline \cr
\mr
No. of Reconstructed Events & 355 & 286 \cr
True Positives & 354 & 285 \cr
False Positives & \hspace{0.35cm}1 & \hspace{0.35cm}1 \cr
Absolute Efficiency & \hspace{0.15cm}18.1\% & \hspace{0.15cm}14.6\% \cr
\br
\end{tabular}
\end{indented}
\end{table}

To compare efficiencies between reconstruction methods, we follow the strategy described in Section \ref{subsc:optimization_baseline} to tune baseline parameters over the test set. In all of the 9805 simulated events, the ML model reconstructed 355 true positive events and only 1 false positive event. The baseline method was manually tuned to match 1 false positive event and subsequently found 285 true positive events. We list this comparison in Table \ref{tb:efficiency_comparison} alongside a measure of absolute efficiency defined as the percentage (no.~of true positives/no.~reconstructable events)$\times 100$ where the denominator is valued as 20\% of 9,805 simulated events. In terms of this measure, the ML model achieves an increase of 3.5\% in absolute efficiency over the baseline method in this study. In terms of total number of reconstructed events alone, the ML-method achieves a relative gain of 24.1\% over the traditional method at the same number of reconstructed false positives. 

As discussed in Section~\ref{subsc:optimization_baseline}, the low-precision matching of baseline and ML false event rates precludes a more precise comparison of true underlying efficiency between models in this study. With higher-statistics simulations producing larger numbers of false events, the baseline false event rate could again be tuned to match the ML false event rate with higher precision, enabling a more direct efficiency comparison in the future. Given that statistical sensitivity is expected to dominate neutrino mass uncertainty in Project 8 (until very large source volumes are employed), and that the statistical variance on the extracted mass scales with the number of reconstructed events as $\sigma_\text{stat}\sim\sqrt{N_\text{tot}}$ \cite{bayesian_paper}, an increase in reconstructed event statistics such as the one shown in this analysis could offer a significant advantage for spectrum reconstruction. 

%%%%%%%%%%%%%%%%%%%% Chapter 7: Conclusion %%%%%%%%%%%%%%%%%%%%
\section{Conclusion and Outlook}
\label{sc:conclusion}

Over the past years, results from the novel CRES technique have shown advances in the field of spectroscopy by exploiting the basic relationship between frequency and energy Eq.~\eqref{eq:cyclotron_frequency} of semi-relativistic particles \cite{Project8:2014ivu,t2_prl,he6_cres}. As represented in the frequency-time plane of a spectrogram, a single CRES event may be multi-faceted, displaying a number of tracks separated by jumps whose starting frequencies must be accurately and precisely extracted in order to faithfully build an energy spectrum. Of particular motivation and application is the goal of the Project 8 experiment: to extract the absolute neutrino mass value with final target sensitivity of 0.04 eV$/c^2$ using the tritium endpoint method. To reach this goal, the experiment must significantly increase its statistics, which is only possible by performing CRES detection over volumes a few orders of magnitude larger than those demonstrated in Phase II \cite{bayesian_paper}. Accordingly, a considerable improvement in event reconstruction efficiency is necessary.

In this work we have presented a ML-based reconstruction method for CRES signals which uses a U-Net CNN architecture in tandem with a SVM to segment and robustly select and build events in the presence of a great abundance of RF noise acting as a background. Besides its relative ease of optimization (training) and featuring a signal-profile agnostic architecture, the ML-method has shown comparable performance in reconstruction accuracy of track parameters and a gain in both absolute ($+3.5\%$) and relative efficiency ($+24.1\%$) in a proof of concept comparison to the baseline approach. The tests of performance of both ML and baseline models were carried out on data produced with an expanded Locust software package, upgraded to simulate realistic CRES-like signals and spectrograms with little computational cost. Future development of this ML-based reconstruction analysis will focus on replacing pre-configured parameters (such as track trimming and first-track selection) for trainable ML submodules and increasing test statistics to provide a precise comparison of true underlying efficiency versus the baseline. The model presented here lays a groundwork for future ML-based analyses in CRES, with the goal of providing a powerful and direct bridge from antenna signal to reconstructed energy spectrum.

%%%%%%%%%%%%%%%%%%%%%%%%%%%%%%%%%%%%%%%%%%%%%%%%%%%%%%%%%%%%%%%%%%%%%%%%%%%%%%%%%%%%%%%%%%%%%%%%%%%%%%%%%%%%%%

\section{Acknowledgements}

This material is based upon work supported by the following sources: the U.S. Department of Energy Office of Science, Office of Nuclear Physics, under Award No.~DE-SC0020433 to Case Western Reserve University (CWRU), under Award No.~DE-SC0011091 to the Massachusetts Institute of Technology (MIT), under Field Work Proposal Number 73006 at the Pacific Northwest National Laboratory (PNNL), a multiprogram national laboratory operated by Battelle for the U.S. Department of Energy under Contract No.~DE-AC05-76RL01830, under Early Career Award No.~DE-SC0019088 to Pennsylvania State University, under Award No.~DE-FG02-97ER41020 to the University of Washington, and under Award No.~DE-SC0012654 to Yale University; the National Science Foundation under Award No.~PHY-2209530 to Indiana University, and under Award No.~PHY-2110569 to MIT; the Cluster of Excellence “Precision Physics, Fundamental Interactions, and Structure of Matter” (PRISMA+ EXC 2118/1) funded by the German Research Foundation (DFG) within the German Excellence Strategy (Project ID 39083149); the Karlsruhe Institute of Technology (KIT) Center Elementary Particle and Astroparticle Physics (KCETA); Laboratory Directed Research and Development (LDRD) 18-ERD-028 and 20-LW-056 at Lawrence Livermore National Laboratory (LLNL), prepared by LLNL under Contract DE-AC52-07NA27344, LLNL-JRNL-840805; the LDRD Program at PNNL; Indiana University; and Yale University.  A portion of the research was performed using the Grace Cluster at the Center for Research Computing of Yale University.  The isotope(s) used in this research were supplied by the United States Department of Energy Office of Science by the Isotope Program in the Office of Nuclear Physics.

\section*{References}
%% Bibliography
\bibliographystyle{iopart-num}
\bibliography{biblio}

\end{document}